# Properties of Farey Sequence and their Applications to Digital Image Processing

Soham Das[1], Kishaloy Halder[1], Sanjoy Pratihar[2], Partha Bhowmick[3]
{88soham, kishaloy.halder, sanjoy.pratihar, bhowmick} @gmail.com
[1] Computer Science & Engg. Deptt., Institute of Engg. & Management, Kolkata, India.
[2] Computer Science & Engg. Deptt., University of Burdwan, Burdwan, India.
[3] Computer Science & Engg. Department, IIT Kharagpur, India.

**Abstract:** *Farey sequence* has been a topic of interest to the mathematicians since the very beginning of last century. With the emergence of various algorithms involving the digital plane in recent times, several interesting works related with the Farey sequence have come up. Our work is related with the problem of searching an arbitrary fraction in a Farey sequence and its relevance to image processing. Given an arbitrary fraction $p/q$ $(0 < p < q)$ and a Farey sequence $\mathcal{F}_n$ of order $n$, we propose a novel algorithm using the *Regula Falsi* method and the concept of *Farey table* to efficiently find the fraction of $\mathcal{F}_n$ closest to $p/q$. All computations are in the integer domain only, which is its added benefit. Some contemporary applications of image processing have also been shown where such concepts can be incorporated. Experimental results have been furnished to demonstrate its efficiency and elegance.

## 1 Introduction

In the year 1816, John Farey invented an amazing procedure to generate proper fractions lying in the interval [0, 1], called the *Farey sequence* [1]. Formally defined, the Farey sequence $\mathcal{F}_n$ of order $n$ is the sequence of simple/irreducible, proper, and positive fractions that have denominators less than or equal to $n$, and are arranged in increasing order of their values. There are several studies and research works related with Farey sequences [2, 3, 4, 5, 7]. The concept is well-known in *theory of fractions* [1, 2, 3] but from the algorithmic point of view very limited work has been done so far. There are two well-known problems in this area. One is the *rank problem*: Given a fraction p/q, find its rank (Fig. 1) in $\mathcal{F}_n$. Another is the *order-statistics problem:* Given two positive integers $n$ and $k$, compute the $k$ th element of $\mathcal{F}_n$. Both are solvable in $O(n \log n)$ time and $O(\sqrt{n})$ space [2, 3]. We show that the rank problem can be solved in $O(1)$ time and $O(n^2)$ space using the *Farey table*, which is an improvement compared to $O(n \log n)$ time [2], although we require $O(n^2)$ space. Our technique avoids floating-point operations as they are expensive and introduce truncations errors. Exhaustive testing and in-depth analyses have been made, and some results have been furnished to demonstrate the performance of the proposed algorithm based on *Regula Falsi* method and *Farey table*.

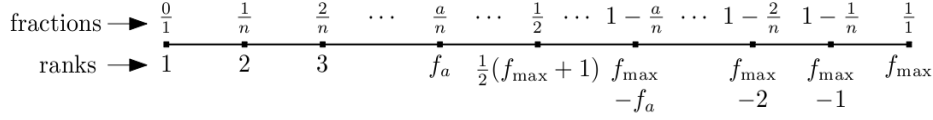

**Fig. 1.** A general *Farey sequence* $\mathcal{F}_n$ having $f_{\max}$ fractions with ranks/indices as $1, 2, \ldots, f_{\max}$.

We show how a Farey sequence $\mathcal{F}_n$ can be used to decide whether three points sorted lexicographically by their *x* and *y* (integer) coordinates, are collinear. This is required to obtain the polygonal approximation of objects present in a digital image. Linearity checking using $\mathcal{F}_n$ involves only addition, comparison, and memory access, but no multiplication. For example, for three points $p_1(i_1, j_1)$, $p_2(i_2, j_2)$, and $p_3(i_3, j_3)$ in succession, conventionally one uses $\Delta(p_1, p_2, p_3) / \max(i_1-i_3, j_1-j_3)$ to decide the deviation of $p_2$ from $p_1p_3$ [2, 10]. However, computation of $\Delta(p_1, p_2, p_3)$ needs multiplication, and hence expensive. Such multiplications are avoided on using $\mathcal{F}_n$.

## 2 Farey Sequence and Farey Table

### 2.1 Farey Sequence

The *Farey sequence* of order *n* is a sequence of irreducible or simple fractions in [0, 1], which have denominators less than or equal to *n*, arranged in ascending order. Each sequence starts with the value $0\,(0/1)$ and ends with the value $1\,(1/1)$. Interestingly, each sequence $\mathcal{F}_n$ can be generated from its preceding sequence $\mathcal{F}_{n-1}$ by inserting the fraction $(a+a')/(b+b')$, called the *mediant* [Sec. 2.2], between each pair of consecutive fractions $a/b$ and $a'/b'$ of $\mathcal{F}_{n-1}$, discarding the fractions whose denominators exceed *n*, as shown in Fig. 2. Note that, barring $\mathcal{F}_1$, each sequence has an odd number of terms and the middle term is always $1/2$.

### 2.2 Properties of the Farey sequence

Some of the properties of the *Farey sequence*, relevant to our work, are discussed below. Their proofs are based on number-theoretic properties and may be seen in [1].

**Property 1:** For any two successive fractions $a/b$ and $c/d$ in $\mathcal{F}_n$, $b+d \geq n+1$, and $bc - ad = 1$ for $ab < cd$.

**Property 2:** Given any real number *x*, there is always a nearby fraction $a/b$ belonging to $\mathcal{F}_n$, such that $|x - a/b| \leq 1/(b(n+1))$.

**Property 3:** For a sufficiently large value of *n*, the number of fractions in $\mathcal{F}_n$ is given by $3n^2/\pi^2 \approx 0.304n^2 = \Theta(n^2)$.

**Property 4:** The *mediant* of two fractions $a/b$ and $c/d$ is defined as $\mu(a/b, c/d) = (a+c)/(b+d)$, which lies in the interval $(a/b, c/d)$. Each fraction in $\mathcal{F}_n$ is the mediant of its two neighbors. In fact, the mediant of any two fractions is contained in the *Farey sequence*, unless the sum of their denominators exceeds *n*.

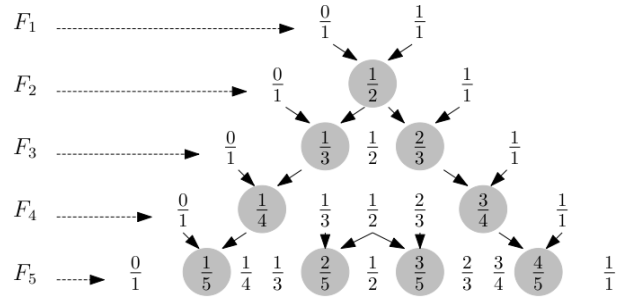

**Fig. 2.** Generation of *Farey sequences* up to order 5 in an iterative way.
(New terms in each sequence are highlighted in gray.)

Apart from the above properties, another striking property is that the rank of a fraction in a *Farey sequence* serves a good measure of its numerical value. A fraction $p/q$ of $\mathcal{F}_n$ has the *rank r* if and only if there exist $r-1$ fractions in $\mathcal{F}_n$, which are smaller than $p/q$ (see Fig. 1). Figure 3 shows a few plots on variation of the numerical values of the fractions with their ranks in *Farey sequences* of different orders. In these plots, barring a few undulations, we can see that the nature of the variation is linear. As the order increases, these undulations are also reduced. Hence, we can say that the ranks are related almost linearly to the numerical values of the fractions. Therefore, given a fraction $x/y$ with rank $k$ in $\mathcal{F}_n$ containing $f_{max}$ fractions, we have $x/y \approx k/f_{max}$. Thus, to compare two fractions we can refer to their ranks in a *Farey sequence* rather than computing and comparing their exact values. In particular, given two fractions and asked to find whether they are close to one another (within a range), we can avoid floating-point operations and can get an estimate by simply looking at the difference of ranks of the given fractions. It is important to note that this difference of ranks will be relative to the order of the concerned Farey sequence.

As mentioned earlier, several methods for generating $\mathcal{F}_n$ are known. Evidently, any such algorithm requires at least quadratic time, since $\mathcal{F}_n$ has $\Theta(n^2)$ elements (Property 3). Hence, the rank problem and the order statistics problem are solvable in $O(n \log n)$ time and $O(\sqrt{n})$ space. In subsequent discussions, we show how we have used a more structured representation of $\mathcal{F}_n$, namely the *Farey table*, $T_n$, to solve the *rank problem* in $O(1)$ time and $O(n^2)$ space.

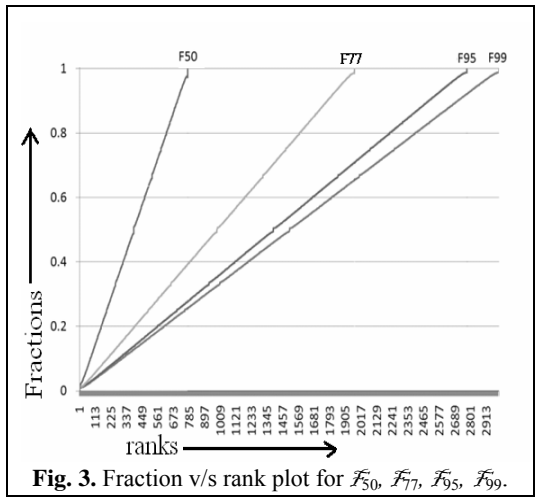

**Fig. 3.** Fraction v/s rank plot for $\mathcal{F}_{50}$, $\mathcal{F}_{77}$, $\mathcal{F}_{95}$, $\mathcal{F}_{99}$.

### 2.3 Farey table

A *Farey table* $T_n$ for $\mathcal{F}_n$ is an $(n+1) \times n$ matrix in which each row $i, 0 \leq i \leq n$, represents a numerator and each column $j, 1 \leq j \leq n$, represents a denominator, and $T_n(i, j)$ contains the rank of the fraction $i/j$ in $\mathcal{F}_n$. For a fraction $i'/j'$ reducible to $i/j$, we have $T_n(i', j') = T_n(i, j)$. As $\mathcal{F}_n$ contains only proper fractions in $[0,1]$, $T_n(i, j)$ is invalid for $i \geq j$. Fig. 4 shows the Farey table $T_4$ of order 4. Note that, if a fraction is in $\mathcal{F}_n$, then we can find its rank from $T_n$ in $O(1)$ time.

**Fig. 4.** $T_4 = \langle p/q : 0 \leq p \leq q \leq 4 \rangle$

### 2.4 Some Properties of Farey Table

The *Farey table* is seen to have some elegant properties. Some of these properties, which are relevant to our work, are as follows.

**Property 5:** The sum of the ranks of two fractions, whose sum is unity, is $f_{\max} + 1$.

*Proof:* For any pair of fractions $i/j$ and $(j-i)/j$ of $\mathcal{F}_n$, their corresponding ranks are of the form $f_i + 1$ and $f_{\max} - f_i$.

**Property 6:** In a Farey table $T_n$, ranks (of fractions from $\mathcal{F}_n$) decrease from left to right along a row, excepting only the first row whose all fractions have rank 1.

*Proof:* Along a row, the numerators of all the fractions are $i$, and the denominators are $i, i+1, i+2, \ldots, n$ (left to right). So for any two fractions $i/j_1$ and $i/j_2$ having $j_1 < j_2$, we get $i/j_1 > i/j_2$, or, $T_n(i, j_1) > T_n(i, j_2)$. Similarly, we have the following property.

**Property 7:** In a Farey table $T_n$, ranks along a column increase downwards.

**Property 8:** Differences between consecutive ranks along a column $j$ in $T_n$ are vertically symmetric with an upward-downward correspondence, i.e.,

$T_n(i+1, j) - T_n(i, j) = T_n(j-i, j) - T_n(j-i-1, j), 0 \leq i \leq j$.

*Proof:* Observe that $\frac{i+1}{j} + \frac{j-i-1}{j} = \frac{i}{j} + \frac{j-i}{j} = 1$. Hence, from Property 5,

$T_n(i+1, j) + T_n(j-i-1, j) = T_n(i, j) + T_n(j-i, j) = f_{\max} + 1$,

$f_{\max}$ being the total number of fractions in $\mathcal{F}_n$.

Notice in Fig. 1 that for each such pair of fractions, the sum of their ranks is $f_{\max} + 1$.

### 2.5 Generation of Farey Table

As explained in Sec. 2.4, we fill up $T_n$ with the ranks of the fractions in $\mathcal{F}_n$. For a fraction $i/j$ in $\mathcal{F}_n$, we generate all other equivalent fractions present in $T_n$ (e.g., $2/4, 3/6, \ldots$ from $1/2$). We fill their cells with the rank of the original irreducible fraction. The algorithm *GenFT* (Fig. 5) generates $T_n$ from $\mathcal{F}_n$ (array $F$). For each $i/j$, we update $T[i][j]$, and all the corresponding equivalent cells. $T_n$ is thus prepared from in $O(n^2)$ time.

```
k ← 1  ▷ k is the rank of a fraction
repeat until (F[k].i ≠ −1)
    l ← 1  ▷  integer to multiply
                numerator & denominator
    T[F[k].i][F[k].j] ← k
    repeat until (F[k].j × l ≤ n)
        T[F[k].i × l][F[k].j × l] ← k
        l ← l + 1
    end loop
end loop
```

**Fig. 5:** Algorithm *GenFT(order: n)*

## 3. Finding the Closest Fraction in $\mathcal{F}_n$

Given any arbitrary proper fraction $p/q$, which may not be present in $\mathcal{F}_n$, we find the closest fraction in $\mathcal{F}_n$ using binary search to reduce the search range of ranks.

### 3.1 Using Binary Search

The range in the sequence, to be searched, can be obtained as follows. Let the search key be $p/q$, $0 \leq p \leq q$. Let $\gcd(\lfloor (pn)/q \rfloor, n) = g_1$ and $\gcd(\lceil (pn)/q \rceil, n) = g_2$. Let $\lfloor (pn)/q \rfloor / n = p_1/q_1$, where $p_1 = \lfloor (pn)/q \rfloor / g_1$ and $q_1 = n/g_1$, s.t. $\gcd(p_1, q_1) = 1$. Also, let $\lceil (pn)/q \rceil / n = p_2/q_2$ ($p_2 = \lceil (pn)/q \rceil / g_2$, $q_2 = n/g_2$), s.t. $\gcd(p_2, q_2) = 1$. It is easy to observe that $p_1/q_1 \leq p/q \leq p_2/q_2$, since $\lfloor (pn)/q \rfloor \leq \lceil (pn)/q \rceil$.

We obtain the ranks of the two fractions, $p_1/q_1$ and $p_2/q_2$, from $T_n$ as $T_n(p_1, q_1) = f_1$ and $T_n(p_2, q_2) = f_2$, respectively. Since the fractions are in ascending order in $\mathcal{F}_n$, we use binary search in $[f_1, f_2]$ to get the fraction closest to $p/q$ in $\mathcal{F}_n$.

**Time complexity:** From Property 4, the maximum rank in $\mathcal{F}_n$ is $3n^2/\pi^2 \approx 0.304n^2 = \Theta(n^2)$. Observe that $\lceil (pn)/q \rceil - \lfloor (pn)/q \rfloor \leq 1$. Thus, the initial length of the search range is $(f_2 - f_1) = f_{\max} \times (m+1-m)/n = f_{\max}/n = O(n^2)/n = O(n)$.

In Figs. 6(a–c), we have shown various cases corresponding to the binary search for a few key fractions in some Farey tables. In these plots, *fraction difference* means the difference of the key $p/q$ from a fraction of $\mathcal{F}_n$. We compute the difference between each fraction of $\mathcal{F}_n$ having rank in $[f_1, f_2]$ and the key fraction $p/q$. These differences are strictly increasing along the range (left to right). They are negative initially and become positive towards right with a transition or *zero-crossing* somewhere in between. Clearly the closest fraction is at the rank nearest to this *zero-crossing*. Hence, we search this *zero-crossing rank* in $O(\log n)$ time using binary search.

## 3.2 Improved Algorithm

There are instances where the zero-crossing is located near one of the boundaries ($f_1$ or $f_2$). In such a case, the binary-search algorithm takes a worst-case time of $O(\log n)$ to find the result. An example is shown in Fig. 6(b). From the three different plots in Fig. 6, we see that the nature is grossly linear. So the idea of *Regula Falsi* method to find the root of a continuous function is used by us as a better alternative. We get two ranks between which this intersecting point lies. The iteration continues depending on the values of the function at these two points. Due to the linearity of the curve, this intersecting point in all cases is found to be very close to the root of the function.

As a result, this new algorithm (given in Fig. 7) finds the root faster.

*Example:* $n = 55$.

Key fraction: $\dfrac{p}{q} = \dfrac{\mathbf{341}}{\mathbf{556}}$.

Closest fraction: $\dfrac{27}{44}$ with rank $f = 577$.

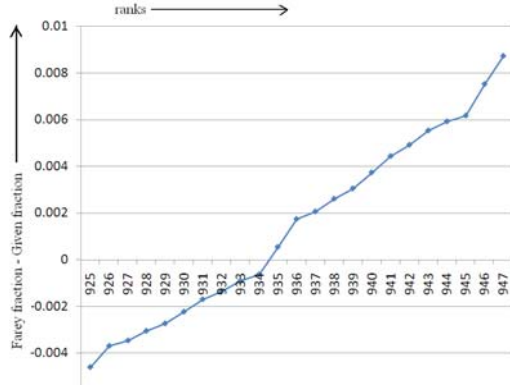

(a) Best case: Searching $p/q = 78/145$ in $\mathcal{F}_{75}$.

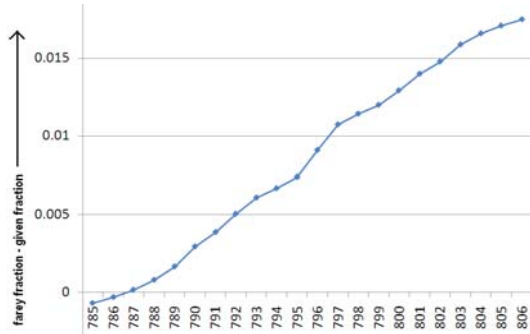

(b) Worst case: Searching $p/q = 375/448$ in $\mathcal{F}_{55}$.

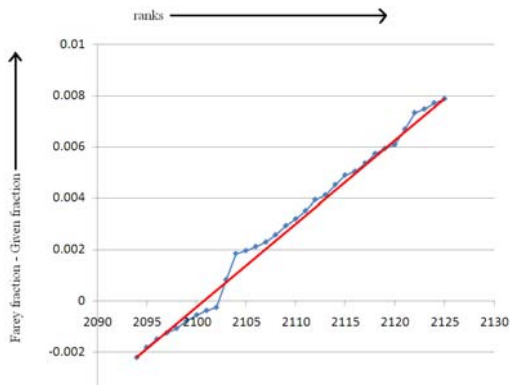

(c) Average case: Searching $1957/2788$ in $\mathcal{F}_{99}$.

**Figure 6:** Instances of best, worst, and average cases for searching three fractions in three Farey tables of orders 75, 55, and 99.

## 3.3 Performance Analysis

To compare the performances of the two algorithms, a set of randomly generated proper fractions in [0,1] is taken and the number of iterations required by these two methods for each of these fractions is reported for various $\mathcal{F}_n$, $50 < n < 400$. The corresponding sets of average number of iterations required by these randomly generated fractions for different values of *n* are plotted against *n* in Fig. 8. Thus, from experimental results also we find that the improved algorithm, *FindClosestFast*, produces significantly better results when compared to the one using binary search as far as the number of iterations is concerned. In particular, the performance of *FindClosestFast* gets even better as *n* increases and the gap between the two performance curves widens farther.

---

**Algorithm** *FindClosestFast* ($\mathcal{F}_n$, $T_n$, $p/q$)

01. compute $p_1/q_1$ and $p_2/q_2$  ▷ Sec. 4.1
02. $f_1 \leftarrow T_n(p_1, q_1)$, $f_2 \leftarrow T_n(p_2, q_2)$  ▷ use $T_n$
03. **if** $f_1 = f_2$ **then return** $p_1/q_1$
04. **if** $f_2 - f_1 = 1$ **then**
05.     **if** $p/q - p_1/q_1 < p_2/q_2 - p/q$ **then return** $p_1/q_1$
06.         **else return** $p_2/q_2$  ▷ else next step
07. compute the differences:
$$y_1 \leftarrow p_1/q_1 - p/q = \frac{p_1 q - q_1 p}{q_1 q}$$
$$y_2 \leftarrow p_2/q_2 - p/q = \frac{p_2 q - q_2 p}{q_2 q}$$
    ▷ *y-coordinates of two points*, $(f_1, y_1)$ *and* $(f_2, y_2)$, *on the line of fraction difference v/s rank.*
08. find the "zero-crossing":  ▷ Sec. 4.1
$$(y - y_1)/(x - f_1) = (y_2 - y_1)/(f_2 - f_1)$$
$$\text{or, } x\big|_{y=0} = (f_1 y_2 - f_2 y_1)/(y_2 - y_1).$$
09. find $p'_1/q'_1$ with rank $f = \lfloor x \rfloor$  ▷ use $\mathcal{F}_n$
10. **if** $p'_1/q'_1 < p/q$ **then** find $p'_2/q'_2$ with rank $\lfloor x+1 \rfloor$
11.     **else** find $p'_2/q'_2$ with rank $\lfloor x-1 \rfloor$  ▷ use $\mathcal{F}_n$
12. **if** $p'_1/q'_1 \leq p/q \leq p'_2/q'_2$ **then**
13.     **if** $p/q - p'_1/q'_1 < p'_2/q'_2 - p/q$ **then return** $p'_1/q'_1$
14.         **else return** $p'_2/q'_2$
15. **else if** $p/q < p'_1/q'_1$ **then**
16.     $p_2/q_2 \leftarrow p'_1/q'_1$, **goto** *Step 07*
17. **else** $p_1/q_1 \leftarrow p'_2/q'_2$, **goto** *Step 07*

**Fig. 7:** The improved algorithm to search an arbitrary fraction $p/q$ ($0 < p < q$) in the Farey sequence, $\mathcal{F}_n$, of an arbitrary order, *n*.

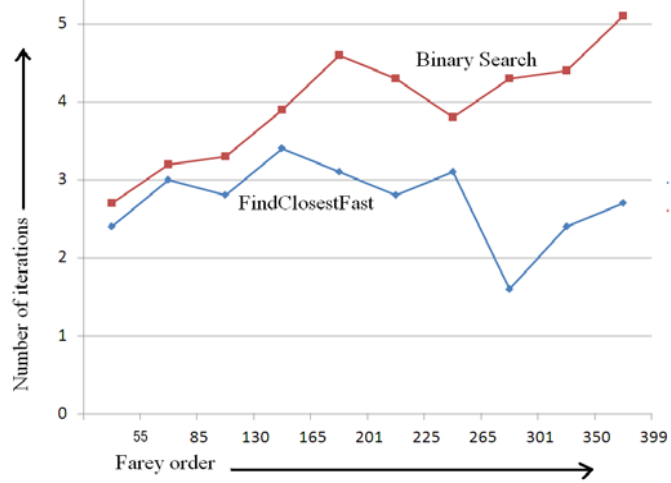

**Fig. 8:** Comparison of the two algorithms based on the average number of iterations v/s order $n$.

## 4. Image-related Applications

### 4.1. Polygonal Approximation

In order to describe an object boundary we use a sequence of straight line segments, which gives a convenient polygonal representation [6, 7, 9]. For an efficient approximation, the successive edges, which are "almost collinear", are merged. In the digital plane where a point means a pixel, the two end-points of a line segment have integer coordinates, and hence we get its slope as $\Delta y/\Delta x$, where $\Delta y$ is the integer difference of $y$-coordinates and $\Delta x$ is that of $x$-coordinates of its end-points. Thus, if the difference of two ranks corresponding to (slopes of) two lines is less than some threshold, $\Delta f$ (depending on order $n$), then two edges, say, AB and BC, are merged into a single edge AC, and AC will be considered further for merging for subsequent lines. As we increase the threshold $\Delta f$, the approximation becomes loose (Fig. 9).

To consider slopes of all possible orientations in the Farey table $T_n$ of order $n$, only the positive proper fractions are included initially. For $n = 10$, initially $T_n$ contains ranks of the slopes or fractions in {0/1 (0/2, 0/3,…, 0/10), 1/10, 1/9,…, 1/3 (2/6, 3/9), …, 7/8, 8/9, 9/10, 1/1 (2/2, 3/3, …, 10/10)}. We take a few reflections (w.r.t. 1/1), interchange the numerator and the denominator, so that we get $\{p/q : -n \leq p, q \leq n\}$. We derive their indices from the fractions in $F_n$, as follows:

$$T_n[i][-j] = 2f_{\max} - T_n[i][j], T_n[-i][-j] = 2f_{\max} - 2 + T_n[i][j], T_n[-i][j]$$
$$= 4f_{\max} - 2 - T_n[i][j].$$

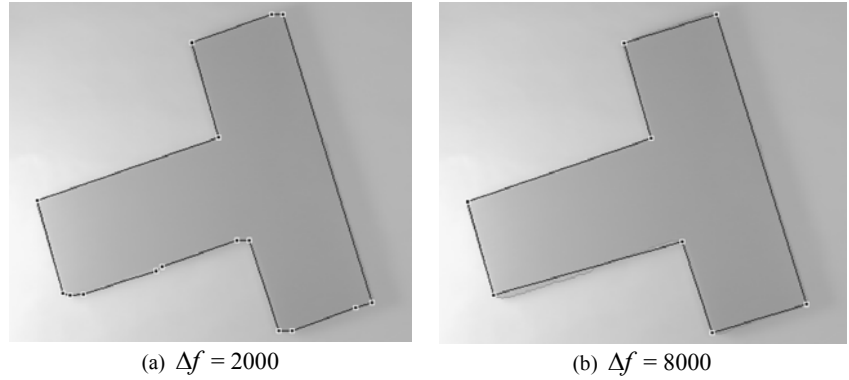

(a) $\Delta f = 2000$          (b) $\Delta f = 8000$

**Fig. 9:** Polygonal approximation for different values of the threshold $\Delta f$ with $n = 200$.

### 4.2 Shape analysis

Shape description of an object is a well-studied, yet ever-engrossing problem. Various shape description and matching techniques are available in the literature, which play significant roles in automated machine recognition of digital objects and patterns. Shape can be given as a numeric sequence, as numeric sequences are easy to represent and analyze. The sequence can be of internal angles of the polygonal boundary, given as the difference of ranks of slopes of successive straight edges. It is almost invariant to rotation; the procedural run-time dealing with shapes is also reduced, as there is no floating-point operation. Using only memory access and subtraction, we can present the descriptor, on which some shape-related algorithms can be applied.

Figure 10 shows the internal angles at the vertices of an approximate polygon. As the object in the gray-scale image rotates, the polygonal approximation gets deviated by an insignificant amount. The interior angles, after re-computation from the re-approximated polygon in the rotated image, are seen to remain almost invariant, which shows the inherent strength of Farey ranks and their differences in capturing shape features. For example, $v_1$ in Fig. 10(a) has a rank-difference $f'_1 = 24650$, which becomes $f'_1 = 24739$ after re-computation. As $n = 200$, we have 97856 entries in $T_n$. Hence, $f'_1 = 24650$ corresponds to $\frac{24650}{97856} \times 360^0 = 90.68^0$ and $f'_1 = 24739$ to $\frac{24739}{97856} \times 360^0 = 91.01^0$; total error is less than $0.50^0$, which is quite small.

### 5. Conclusion

This paper shows that the ranks of fractions in the *Farey sequence* can be used to provide a useful estimation of their relative values. Searching a fraction in the *Farey sequence* has been improved by using the *Farey table*. Algorithms to find a fraction closest to any arbitrary fraction in a given *Farey sequence* are also presented. All the algorithms are devoid of floating-point operations, and thus save time to perform their underlying functions. The *Farey table* can have wide applications in digital image processing and shape analysis, as shown in this paper. It also puts forward some important issues, such as the problem of compressing the table by removing some columns such that the maximal difference of ranks in the contiguous columns is minimized. We are presently working on this, and relevant outcomes will be reported in near future.

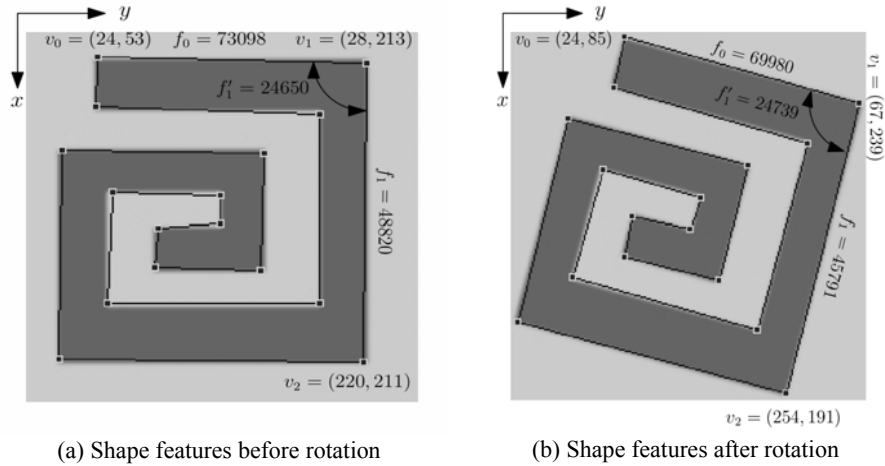

(a) Shape features before rotation   (b) Shape features after rotation

**Figure 10:** Invariance of shape features (differences of ranks in $T_n$) with rotation.